\documentclass[preprint,12pt]{elsarticle}

 \usepackage{subfigure}
 \usepackage{graphics}
\usepackage{graphicx}
 \usepackage{epsfig}
 \usepackage{epstopdf}
\usepackage{amssymb}
\usepackage{tabularx}
\usepackage{color}

\usepackage[abs]{overpic}

\usepackage[colorlinks, allcolors=blue]{hyperref} 

\renewcommand\Re{\mathit{Re}}  
\newcommand\Ra{\mathit{Ra}}
\newcommand\Cn{\mathit{Cn}}
\newcommand\Ca{\mathit{Ca}}
\newcommand\We{\mathit{We}}
\newcommand\Fr{\mathit{Fr}}
\newcommand\Bo{\mathit{Bo}}
\renewcommand\Pr{\mathit{Pr}} 
\newcommand\Pe{\mathit{Pe}}  

\journal{Journal of Computational Physics}

\begin{document}

\begin{frontmatter}

\title{An efficient phase-field method for turbulent multiphase flows}

\author[label1]{Hao-Ran Liu\corref{cor1}}

\cortext[cor1]{h.liu-2@utwente.nl}

\author[label1]{Chong Shen Ng}

\author[label1]{Kai Leong Chong}

\author[label1,label2]{Detlef Lohse}

\author[label3,label4,label1]{Roberto Verzicco\corref{cor2}}

\cortext[cor2]{verzicco@uniroma2.it}

\address[label1]{Physics of Fluids Group and Max Planck Center for Complex Fluid Dynamics, MESA+ Institute and J. M. Burgers Centre for Fluid Dynamics, University of Twente,\\
P.O. Box 217, 7500AE Enschede, The Netherlands}
\address[label2]{Max Planck Institute for Dynamics and Self-Organization, Am Fassberg 17, 37077 G\"ottingen, Germany}
\address[label3]{Dipartimento di Ingegneria Industriale, University of Rome ``Tor Vergata", Via del Politecnico 1, Roma 00133, Italy}
\address[label4]{Gran Sasso Science Institute - Viale F. Crispi, 7 67100 L’Aquila, Italy}





\begin{abstract}
With the aim of efficiently simulating three-dimensional multiphase turbulent flows with a phase-field method, we propose a new discretization scheme for the biharmonic term (the 4th-order derivative term) of the Cahn-Hilliard equation.
This novel scheme can significantly reduce the computational cost while retaining the same accuracy as the original procedure. Our phase-field method is built on top of a direct numerical simulation solver, named AFiD (\href{https://github.com/PhysicsofFluids/AFiD}{www.afid.eu}) and open-sourced by our research group. It relies on a pencil distributed parallel strategy and a FFT-based Poisson solver. To deal with large density ratios between the two phases, a pressure split method \cite{jcp14} has been applied to the Poisson solver. To further reduce computational costs, we implement a multiple-resolution algorithm which decouples the discretizations for the Navier-Stokes equations and the scalar equation: while a stretched wall-resolving grid is used for the Navier-Stokes equations, for the Cahn-Hilliard equation we use a fine uniform mesh. The present method shows excellent computational performance for large-scale computation: on meshes up to 8 billion nodes and 3072 CPU cores, a multiphase flow needs only slightly less than $1.5$ times the CPU time of the single-phase flow solver on the same grid. The present method is validated by comparing the results to previous studies for the cases of drop deformation in shear flow, including the convergence test with mesh refinement, and breakup of a rising buoyant bubble with density ratio up to $1000$. Finally, we simulate the breakup of a big drop and the coalescence of $O(10^3)$ drops in turbulent Rayleigh-B\'enard convection at a Rayleigh number of $10^8$, observing good agreement with theoretical results.

\end{abstract}


\end{frontmatter}

\section{Introduction}
\label{sec-intro}

Turbulent multiphase flows are ubiquitous in nature and technology. Examples are raindrops \citep{rain, zaleski}, ocean waves \cite{spray}, fuel sprays \cite{fuel}, and the transmission of virus-laden droplets during respiratory events \cite{covid,covid-steven,covid-cs}, just to name a few. In order to gain deeper insights into their complex and rich behavior, efficient, high-fidelity computations are crucial. For turbulent multiphase flows, direct numerical simulations (DNSs) present far greater challenges than for single-phase flows \cite{rev-tmf}. The reasons are the much finer length-scales and faster time-scales induced by the existence of the second phase, especially when the deformable interfaces between the fluids break up or coalesce.

To-date, many numerical methods have been developed, such as phase field methods (also known as diffuse interface methods) \cite{soldati1, soldati2, breakup}, volume of fluid methods \cite{pop16, luka19jfm}, level set methods \cite{ls-tmf}, and front tracking \cite{ft-tmf}, Lattice-Boltzmann \cite{LB19JFM}, and immersed boundary \cite{roberto-jcp, cs20} methods. Among them, the phase-field method is an approach in which a scalar (volume fraction of one fluid) is tracked by the Cahn-Hilliard equation and the sharp fluid-fluid interface is replaced by a narrowly mixed layer \cite{jacqmin99}. 
In the past decade, application of the phase-field method has been increasingly appealing because of its versatility. For example, the method has been applied to the simulation of turbulent flows \cite{soldati1, soldati2, soldati4, breakup}, flows with moving contact lines \cite{liu15,sui14,ding07jfm,zy}, fluid-structure interaction \cite{chen1,liu17,chen,liu20}, melting flows \cite{melt1,melt2,melt3}, ternary flows \cite{liu18}, and even brittle fracture simulation \cite{brittle}.

In the phase-field method, two immiscible phases are represented by their volume fractions $C$ and $1-C$, respectively. The spatial distribution of $C$ is determined by the Cahn-Hilliard equation \cite{pfm96,jacqmin99,ding07jcp}:
\begin{equation} \frac {\partial C} {\partial t} + \nabla \cdot ({\bf u} C) = M \nabla^2  \left[a_1\nabla^2 C+a_2\psi'(C)\right].
\label{eq-ch-d}
\end{equation} 
The quantity in square brackets is the chemical potential defined by the variation of free energy with respect to $C$. It includes an excess free energy term (the first term), and a bulk energy term (the second term) with $\psi=1/4\,C^2(C-1)^2$ being the simplest non-singular form that has two equal energy minima, namely at $C=0$ and $1$ \cite{pfm96,jacqmin99,ding07jcp}. Physically, $\psi$ represents the bulk energy density due to the inhomogeneous distribution of volume fraction in the interfacial region. We will give more technical details in Section \ref{sec-ch}. In Eq.~(\ref{eq-ch-d}), the Laplacian of the first term on the right hand side is biharmonic, i.e. it contains fourth-order derivatives.

State-of-the-art solvers for the standard single phase flow Navier-Stokes equation is highly efficient and well-studied for turbulent flows. This is because the typical algorithm to solve the computationally demanding Poisson equation--a necessary step for enforcing incompressibility--is based on fast Fourier transforms (FFTs) \cite{fft1,fft2}, as described in Ref.~\cite{cf15}. In Refs.~\cite{jcp12} and \cite{jcp14}, the FFTs is extended to multiphase flows by employing a split method, meaning the variable-coefficient pressure-gradient term is spilt into an implicit constant term and an explicit variable term. As a result, the Poisson equation can be solved up to $40$ times faster than without the split method \cite{jcp14}.

\begin{figure}
\centering
\includegraphics[width=0.45\linewidth]{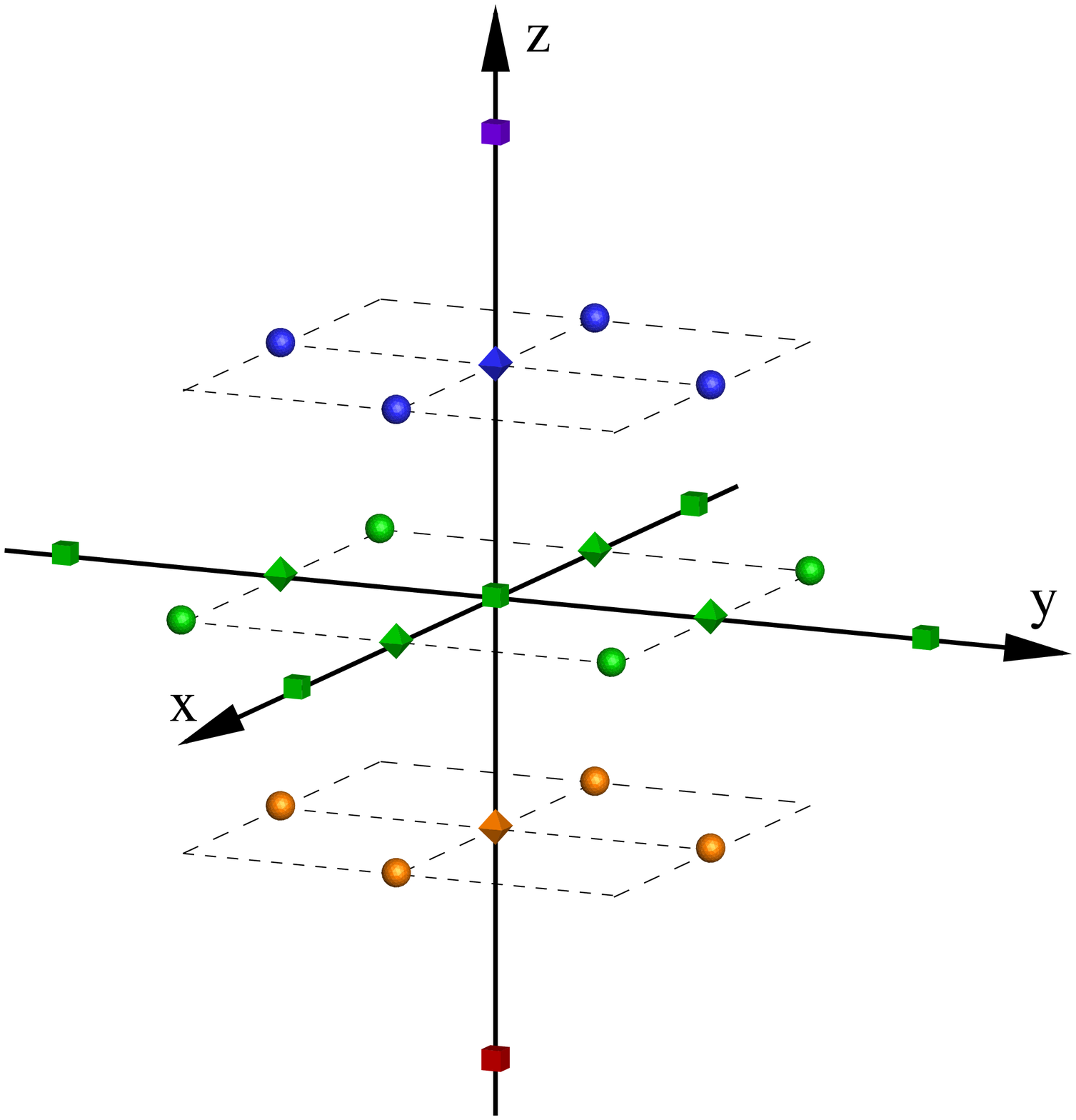}
\hspace{0.05\linewidth}
\includegraphics[width=0.45\linewidth]{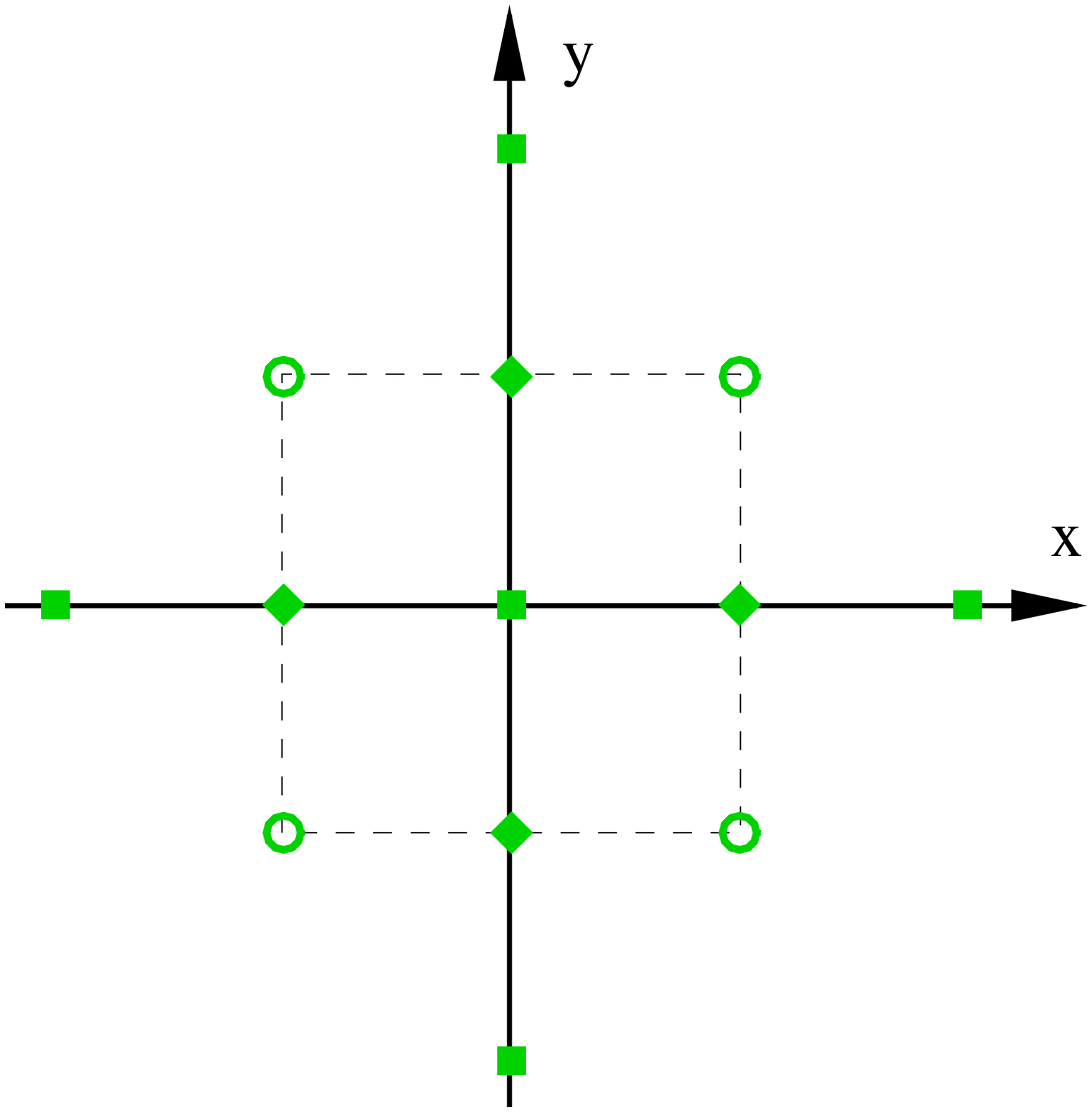}
\caption{\label{fig-dis}  (a) $25$ points used to discretize the biharmonic terms in the Cahn-Hilliard equation (\ref{eq-ch}) in the scheme of Eq.~(\ref{eq-disbi}). Symbols in different colors represents the points at different $z$ plane. In the new discretization scheme of Eq.~(\ref{eq-new}), the spherical points are replaced by the cubic ones. (b) Two dimensional situation of the discretization of the biharmonic terms. The circles are replaced by the square ones.}
\end{figure}

However, with the application of FFTs in multiphase flows, the computational cost of the biharmonic term becomes the new bottleneck for the phase-field method. The reason for this is that the common solution technique for the biharmonic term in the phase-field method involves an implicit solution that requires $25$ grid points for a second-order spatial discretization, see Fig.~\ref{fig-dis}(a) (details in Section \ref{sec-bi}). Therefore, in this study, we will in particular focus on an optimal discretization of the biharmonic term. We propose a novel discretization scheme for the biharmonic term in the phase-field method to couple with the approximate-factorization method, which is an efficient way to implicitly solve the hyperbolic systems \cite{axayaz} and easily parallelize it. We will implement the phase-field method \cite{ding07jcp} with this novel scheme into our open-source DNS package AFiD (\href{https://github.com/PhysicsofFluids/AFiD}{www.afid.eu}) \cite{jcp96,cf15}, which is a second-order finite difference solver that has been well-validated in many studies of turbulent flows \cite{zhu-tc,shan,qi3}. AFiD is highly-parallelized with a pencil distributed strategy \cite{cf15,gpu}, and includes an FFT-based Poisson solver \cite{jcp96}. In addition, we will apply a split method \cite{jcp12,jcp14} to the pressure solver to deal with large density differences between the two phases. 

To validate the present approach, we simulated cases of drop deformation in a shear flow and of a rising buoyant bubble. Our results are compared to previous studies and are further assessed using a grid convergence study. Finally, we simulated the case of a breakup of one big drop as well as the coalescence of $O(10^3)$ drops in turbulent Rayleigh-B\'enard convection, and show the good performance of the present approach for large-scale computation.

The paper is organized as follows. The governing equations are introduced in Section \ref{sec-ge}. Then we address the numerical methodology in Section \ref{sec-num}. In Section \ref{sec-case}, we simulate several test cases to validate our approach and show its ability to deal with turbulent multiphase flows in large-scale computation. We conclude our study in Section \ref{sec-con}.

\section{Governing Equations}
\label{sec-ge}

\subsection{Cahn-Hilliard (CH) equation}
\label{sec-ch}
Turbulent flows with two incompressible immiscible fluids are investigated here. We use the phase-field method \cite{jacqmin00,ding07jcp} to capture the interface between two fluids. Here, the sharp interface is modeled by a diffused one with finite thickness, and represented by contours of the volume fraction $C$ of fluid $1$, and thus the volume fraction of fluid $2$ is $1-C$. The evolution of the volume fraction $C$ is governed by the Cahn-Hilliard equation,

\begin{equation}
\begin{array}{ll}
\displaystyle \frac {\partial C} {\partial t} + \nabla \cdot ({\bf u} C)&\displaystyle=\frac{1}{\Pe}\left[-\Cn ^{2} \nabla^4 C+\nabla^2  \left(C^{3} - 1.5 C^{2}+ 0.5 C\right)\right],
\end{array}
\label{eq-ch}
\end{equation} 
where $\bf u$ is the flow velocity. We choose the P\'eclet number (the ratio of advection and diffusion) and the Cahn number (a dimensionless measure of the thickness of diffuse interface) the same as in Ref.~\cite{liu15}, i.e. $\Pe=0.9\Cn$ and $\Cn=0.75h/L$ with $h$ and $L$ the uniform mesh size and the characteristic length, respectively.

To enforce mass conservation, the correction method proposed by \cite{shu} is used. This correction method resembles that of Ref.~\cite{soldati3} and exhibits good performance (see Section \ref{sec-rb}).

\subsection{Navier-Stokes (NS) equations}
\label{sec-ns}

The fluid motion is governed by the momentum and continuity equations,
\begin{equation}
\rho\left(\frac {\partial {\bf u}} {\partial t} + {\bf u} \cdot \nabla {\bf u}\right)= - \nabla P + \frac{1}{\Re} \nabla \cdot \mu (\nabla {\bf u}+ \nabla {\bf u}^{T}) + \frac{\bf F_{st}}{\We}+ {\bf G},
\label{eq-ns}
\end{equation}

\begin{equation}
\nabla \cdot {\bf u}= 0,
\label{eq-con}
\end{equation}
which have been made dimensionless using the properties of fluid $1$. Here, $\bf u$ is the velocity and $P$ the pressure. $\rho$ and $\mu$ are the density and the dynamic viscosity, respectively, which are both functions of $C$ defined as,
\begin{equation}
\rho =C + \lambda_\rho(1-C),
\label{eq-rho}
\end{equation}

\begin{equation}
\mu =C + \lambda_\mu(1-C),
\label{eq-mu}
\end{equation}
where $\lambda_\rho=\rho_2/\rho_1$ and $\lambda_\mu=\mu_2/\mu_1$ are the ratio of the densities and viscosities of the two phases (denoted by the subscript), respectively. The surface tension force ${\bf F}_{st}$ is computed as in \cite{ding07jcp}, 
\begin{equation}
 {\bf F}_{st} =6\sqrt{2}\phi \nabla C / \Cn. 
\label{eq-fst}
\end{equation}
In Eq.~(\ref{eq-ns}), the gravity force is ${\bf G}=-\rho/\Fr \, {\bf j}$ with $\bf j$ being the vertical direction. The dimensionless numbers controlling the problems are thus the Reynolds number $\Re=\rho_1UL/\mu_1$, the Weber number $\We=\rho_1 U^2 L/\sigma$, and the Froude number $\Fr=U^2/(gL)$, where $\sigma$ the surface tension coefficient, $g$ the gravity acceleration, and $U$ is the characteristic velocity. 

\section{Numerical method}
\label{sec-num}
We use staggered meshes and solve the CH equation on the uniform mesh with size $h$ for all three directions and the NS equations on the stretched mesh: the procedure for the coupling of the two meshes (uniform and stretched) is based on that reported in \cite{rodolfo} and it is described in Section \ref{sec-mesh}. A low-storage third-order Runge-Kutta method \cite{rk} is used to temporally advanced all the equations. The biharmonic term in Eq.~(\ref{eq-ch}), viscosity term in Eq.~(\ref{eq-ns}), and diffusion term in Eq.~(\ref{eq-t}) are implicitly solved by the Crank-Nicolson scheme, while the other terms are solved explicitly. In spatial discretization, central second-order accurate finite-difference schemes are used for all terms (details can be found in \cite{jcp96,ding07jcp}), except for two: one is the advection term of volume fraction $C$ in CH equation (\ref{eq-ch}), which is solved by fifth-order WENO scheme \cite{ding07jcp}, and the other is the biharmonic term which is solved by a novel scheme proposed in Section \ref{sec-bi}.

\subsection{Discretization of biharmonic term in CH equation}
\label{sec-bi}

To accurately advance the CH equation (\ref{eq-ch}) with a large time step, we should implicitly solve the biharmonic term $\Cn^2\nabla^4 C$ at the right-hand side of Eq.~(\ref{eq-ch}). At the same time, its discretization scheme should retain the same order of error as the term $\nabla^2(C^3-1.5C^2+0.5C)$, which is also at the right-hand side of Eq.~(\ref{eq-ch}) and discretized by central second-order finite-difference schemes of $O(h^2/L^2)$.

Typically, the biharmonic term is discretized according to Fig.~\ref{fig-dis}(b) (we restrict the expression to a 2D case for the ease of representation),
\begin{equation}
\begin{array}{rl}
(\Cn^2\nabla^4 C)_{i,j}=&\displaystyle \left(\frac{0.75h}{L}\right)^2\left(\frac{\partial^4 C}{\partial x^4}+\frac{\partial^4 C}{\partial y^4}+\frac{2\partial^4 C}{\partial x^2 \partial y^2}\right)_{i,j}\\
\\
=&\displaystyle \left(\frac{0.75h}{L}\right)^2\left(\frac{L}{h}\right)^4(C_{i-2,j}  -8 C_{i-1,j}  +20 C_{i,j} -8 C_{i+1,j}  + C_{i+2,j} \\
& +C_{i,j+2}-8 C_{i,j+1}-8C_{i,j-1}+C_{i,j-2} \\
&  +2C_{i-1,j+1} +2 C_{i+1,j+1}+2C_{i-1,j-1}+2C_{i+1,j-1})\\
\\
& +O(h^4/L^4).
\end{array}
\label{eq-disbi}
\end{equation}
When we implicitly solve this expression, the presence of mixed partial derivatives poses challenges for computational cost and code parallelisation.

To circumvent the use of mixed partial derivatives when solving Eq.~(\ref{eq-disbi}), we propose a new discretization scheme, which is shown in Eqs.~(\ref{eq-new}), (\ref{eq-a2d}) and (\ref{eq-a3d}). Thus, we can split this discretization into two one-dimensional parts $A_x C$ with $C_{i+m,j}$ and $A_y C$ with $C_{i,j+n}$,
\begin{equation}
\Cn^2\nabla^4 C=(A_x+A_y)C,
\label{eq-divbi}
\end{equation}
which means that only the points on the axes remain (Fig.~\ref{fig-dis}b). Then, we can use the approximate-factorization method (described at the end of this section) to efficiently solve $\Cn^2\nabla^4 C$ implicitly.

Our main idea is replacing $C_{i\pm1,j\pm1}$ in Eq.~(\ref{eq-disbi}) with $C_{i+m,j}$ and $C_{i,j+n}$ (Fig.~\ref{fig-dis}b), where $m$ and $n=-2,-1,0,1,2$. The replacement is justified based on the Taylor series expansions,
\begin{equation}
\left\{\begin{array}{lr}
C_{i+m,j}=  C_{i,j} \quad + m (h/L) C'_x \quad+ m^2 (h/L)^2 C''_x/2 & +\,m^3 (h/L)^3 C'''_x/6\quad+O(h^4/L^4),\\ 
& m=-2,-1,0,1,2;\\
\\
C_{i,j+n}=\ C_{i,j}\quad + n (h/L) C'_y\quad +n^2 (h/L)^2 C''_y/2 & +\,n^3 (h/L)^3 C'''_y/6\quad+O(h^4/L^4),\\
& n=-2,-1,0,1,2;\\
\\
C_{i+m,j+n}=C_{i,j}\ +\sqrt{2}m (h/L) C'_s\ +m^2(h/L)^2 C''_s & +\sqrt{2}m^3 (h/L)^3 C'''_s/3\ +O(h^4/L^4),\\ & (m,n)=(-1,1), (1,-1);\\
\\
C_{i+m,j+n}=C_{i,j}\ +\sqrt{2}m (h/L) C'_\tau\ +m^2(h/L)^2 C''_\tau & +\sqrt{2}m^3 (h/L)^3 C'''_\tau/3\ +O(h^4/L^4),\\
& (m,n)=(1,1), (-1,-1);\\
\end{array}\right.
\label{eq-taylor}
\end{equation}
where we define $C'_e=(\partial C/\partial e)_{i,j}$ with $e=x$, $y$, $s$ and $\tau$, so do $C''_e$ and $C'''_e$. The directions $x$ and $y$ are the perpendicular axis directions in Cartesian coordinates, and the directions $s$ and $\tau$ are obtained by rotating $x$ and $y$ by $45^\circ$. Since the Laplacian operator is rotational invariant, we have
\begin{equation}
    \nabla^2 C = C''_s+C''_\tau=C''_x+C''_y,
    \label{eq-c2}
\end{equation} 
so we have the relations,
\begin{equation}
\begin{array}{rl}
&C_{i+1,j+1}+C_{i+1,j-1}+C_{i-1,j+1}+C_{i-1,j-1}\\
=&4C_{i,j}+2(h/L)^2(C''_s+C''_\tau)+O(h^4/L^4)\\
=&4C_{i,j}+2(h/L)^2(C''_x+C''_y)+O(h^4/L^4)\\
=&2C_{i,j}+0.5\{[C_{i,j}+2^2 (h/L)^2 C''_x/2+O(h^4/L^4)]+[C_{i,j}+(-2)^2 (h/L)^2 C''_x/2\\&+O(h^4/L^4)]\}\\&+0.5\{[C_{i,j}+2^2 (h/L)^2 C''_y/2+O(h^4/L^4)]+[C_{i,j}+(-2)^2 (h/L)^2 C''_y/2\\&+O(h^4/L^4)]\}\\
=&0.5(2C_{i,j}+C_{i+2,j}+C_{i-2,j})+0.5(2C_{i,j}+C_{i,j+2}+C_{i,j-2})+O(h^4/L^4),
\end{array}
\label{eq-trans1}
\end{equation}
where the first and third-order derivatives are eliminated since the points are symmetrical about $(i,j)$. Thus, $C_{i\pm1,j\pm1}$ can be replaced by $C_{i+m,j}$ and $C_{i,j+n}$ as shown in Fig.~\ref{fig-dis}(b).

Substituting Eq.~(\ref{eq-trans1}) into  Eq.~(\ref{eq-disbi}), we get the new discretization scheme,
\begin{equation}
\begin{array}{rl}
(\Cn^2\nabla^4 C)_{i,j}=&\displaystyle \left(\frac{0.75h}{L}\right)^2\left(\frac{L}{h}\right)^4(C_{i-2,j}  -8 C_{i-1,j}  +20 C_{i,j} -8 C_{i+1,j}  + C_{i+2,j} \\
&  +C_{i,j+2}-8 C_{i,j+1} -8C_{i,j-1}+C_{i,j-2}\\
& +2C_{i,j}+C_{i+2,j}+C_{i-2,j}+2C_{i,j}+C_{i,j+2}+C_{i,j-2})\\
\\
& +O(h^2/L^2)\\
\\
=&\displaystyle \left(\frac{0.75h}{L}\right)^2\left(\frac{L}{h}\right)^4(2C_{i-2,j}  -8 C_{i-1,j}  +10 C_{i,j} -8 C_{i+1,j}  + 2C_{i+2,j} \\
&  +2C_{i,j+2}-8 C_{i,j+1}+10 C_{i,j} -8C_{i,j-1}+2C_{i,j-2}) \\
\\
& +O(h^2/L^2),
\end{array}
\label{eq-new}
\end{equation}
where the error $O(h^2/L^2)$ is of the same order as the term $\nabla^2(C^3-1.5C^2+0.5C)$ at the right-hand side of Eq.~(\ref{eq-ch}). Comparing  Eq.~(\ref{eq-new}) and  Eq.~(\ref{eq-divbi}), we get the following pentadiagonal matrix,

\begin{equation}
A_x=A_y=\displaystyle \left(\frac{0.75L}{h}\right)^2
\left[\begin{array}{cccccccc}
\cdots &&&\cdots&&&&\cdots\\
2&-8&12&-8&2 &&&0\\
0&\ddots&\ddots&\ddots&\ddots&\ddots &&\vdots\\
\vdots&&\ddots&\ddots&\ddots&\ddots&\ddots &0\\
0&&    &2&-8&12&-8&2  \\
\cdots &&&&\cdots&&&\cdots\\
\end{array}\right],
\label{eq-a2d}
\end{equation}
for 2D, where the values in the first and last rows are determined by boundary conditions. Now, with the convenient form of Eq.~(\ref{eq-a2d}), the approximate-factorization method can be employed to solve the biharmonic term implicitly. The same idea can be directly extended to three-dimensions, and the points used in the mixed partial derivatives are replaced as shown in Fig.~\ref{fig-dis}(a). Thus, we get the operators,
\begin{equation}
A_x=A_y=A_z=\displaystyle \left(\frac{0.75L}{h}\right)^2
\left[\begin{array}{cccccccc}
\cdots &&&\cdots&&&&\cdots\\
4&-16&24&-16&4&&&0\\
0&\ddots&\ddots&\ddots&\ddots&\ddots &&\vdots\\
\vdots&&\ddots&\ddots&\ddots&\ddots&\ddots &0\\
0&&    &4&-16&24&-16&4  \\
\cdots &&&&\cdots&&&\cdots\\
\end{array}\right],
\label{eq-a3d}
\end{equation}
for 3D.

With $A_x$, $A_y$ and $A_z$, we can use the approximate-factorization method \cite{axayaz,jcp96} to efficiently solve the following equation with the known $q^l$ from the previous time step and unknown $q^{l+1}$ for the next time step,
\begin{equation}\label{eq-example}
\frac{q^{l+1}-q^l}{\delta t} = E+\beta(A_x +A_y +A_z)\frac{q^{l+1}+q^l}{2},
\end{equation}
where $E$ represents the terms calculated explicitly, $\beta$ is the constant coefficient, $A_x$, $A_y$ and $A_z$ are discretization operators, and $(q^{l+1}+q^l)/2$ originates from the Crank-Nicolson scheme.

Eq.~(\ref{eq-example}) can be rewritten as,
\begin{equation}\label{eq-re}
\left[1-\frac{\delta t \beta}{2}(A_x +A_y +A_z)\right](q^{l+1}-q^l) = \delta t E+ \delta t \beta(A_x +A_y +A_z) q^l.
\end{equation}
Then we factorize the operators on the left,
\begin{equation}\label{eq-fac}
\left[1-\frac{\delta t \beta}{2}(A_x +A_y +A_z)\right] = \left(1-\frac{\delta t \beta}{2}A_x\right) \left(1-\frac{\delta t \beta}{2}A_y\right) \left(1-\frac{\delta t \beta}{2}A_z\right)+O(\delta t^2 \beta^2).
\end{equation}
After factorization, the computation only requires inversions of separate tridiagonal matrices rather than the inversion of a large sparse matrix, which leads to a significant reduction in computation cost and memory \cite{axayaz,jcp96}. Then, Eq.~(\ref{eq-re}) can be solved by the following steps,
\begin{equation}
\label{eq-fac-1}
\left(1- \frac{\delta t \beta}{2}A_x\right)\delta q^* = \delta t E+ \delta t \beta(A_x +A_y +A_z) q^l,
\end{equation}
\begin{equation}
\label{eq-fac-2}
\left(1- \frac{\delta t \beta}{2}A_y\right)\delta q^{**} = \delta q^*,
\end{equation}
\begin{equation}
\label{eq-fac-3}
\left(1- \frac{\delta t \beta}{2}A_z\right)(q^{l+1}-q^l) = \delta q^{**},
\end{equation}
where the superscript $*$ represents the intermediate parameter. In Eqs.~(\ref{eq-fac-1}), (\ref{eq-fac-2}) and (\ref{eq-fac-3}), the inversion of matrix will be extremely cheap when we carefully choose $A_x$, $A_y$ and $A_z$, respectively, provided they only involve the points in one dimension.

\subsection{FFT-based solver with a split method for Poisson equation with large density contrast}
\label{sec-fft}

The NS equation (\ref{eq-ns}) is solved here by a projection method,
\begin{equation}
    \frac{{\bf u}^{l+1}-{\bf u}^{*}}{\delta t}=-\frac{1}{\rho^{l+1}}\nabla P^{l+1},
\label{eq-pro}
\end{equation}
where ${\bf u}^*$ is an intermediate velocity field calculated from Eq.~(\ref{eq-ns}) without the pressure term. Considering $\nabla \cdot {\bf u}^{l+1}=0$, we have,
\begin{equation}
\nabla \cdot \left(\frac{1}{\rho^{l+1}}\nabla P^{l+1}\right)=\frac{1}{\delta t}\nabla \cdot {\bf u}^{*}.
\label{eq-poi}
\end{equation}
To solve this Poisson equation with large density variations, we use the split method proposed by \cite{jcp14} to apply fast Poisson solver to  Eq.~(\ref{eq-poi}). In the split method \cite{jcp14}, the Poisson equation (\ref{eq-poi}) with the variable coefficient $1/\rho^{l+1}$ is split into an implicit constant density part and an explicit variable part,
\begin{equation}
    \frac{1}{\rho^{l+1}}\nabla P^{l+1}=\frac{1}{\rho_2}\nabla P^{l+1}+\left(\frac{1}{\rho^{l+1}}-\frac{1}{\rho_2}\right)\nabla (2P^l-P^{l-1}),
\label{eq-split}
\end{equation}
where we define $\rho_2 \le \rho_1$. Substitute  Eq.~(\ref{eq-split}) into  Eq.~(\ref{eq-poi}), 

\begin{equation}
    \nabla^2 P^{l+1}=\nabla \cdot \left[\left(1-\frac{\rho_2}{\rho^{l+1}}\right)\nabla (2P^l-P^{l-1})\right]+\frac{\rho_2}{\delta t}\nabla \cdot {\bf u}^{*}.
\label{eq-newpoi}
\end{equation}
Then, a standard fast Poisson solver can be used here. After getting $P^{l+1}$, the velocity field is updated as,
\begin{equation}
    {\bf u}^{l+1}={\bf u}^{*}-\delta t\left[\frac{1}{\rho_2}\nabla P^{l+1}+\left(\frac{1}{\rho^{l+1}}-\frac{1}{\rho_2}\right)\nabla (2P^l-P^{l-1})\right].
\label{eq-u}
\end{equation}

\subsection{Pencil distributed parallel strategy}
\label{sec-para}
The parallel method in the present approach is a pencil distributed parallel strategy (details in \cite{cf15}). Here, the computational domain is split in two dimensions and this strategy allows us to use more CPU cores for large-scale computation, such as $70$ billion points with $64K$ cores as reported in \cite{cf15}. The other advantage is that this strategy is well coupled with the approximate-factorization method to implicitly solve the equations. The high performance of this parallel method has been extensively validated in \cite{cf15} and \cite{gpu}. Moreover, it has already been used in many studies of turbulent flows in large-scale simulations \cite{richard,zhu-tc,shan,zhu}.

\subsection{Multi-resolution meshes for $C$ and $\bf u$}
\label{sec-mesh}
One feature of our method is that the volume fraction field $C$ can be integrated on a refined uniform mesh, even if the momentum field $\bf u$ is integrated on a non-uniform mesh. For the $C$ field, a uniform mesh is a recommended choice. The reasons for this are as follows: The computation of surface tension force is key to simulate multiphase flows. To ensure the truncation error of surface tension in space is of the same order in all directions, uniform mesh spacing in each direction is necessary near the interface. Furthermore, considering the drops in turbulent flows is likely to break up into smaller sized drops and distribute throughout the domain, the use of a uniform mesh can easily handle the spatially dispersed drops. Therefore, the uniform mesh is a good choice for $C$ field.

On the other hand, in wall-bounded turbulence, the resolution requirements of the $\bf u$ field are more restrictive at the walls, where very thin kinematic boundary layers need to be resolved. The same strict requirements apply for Rayleigh--B\'{e}nard convection, where a large number of near-wall nodes are required to resolve thin thermal boundary layers \cite{olga}. Therefore, a stretched non-uniform mesh is a good choice for resolving $\bf u$ or the temperature field. This multi-resolution treatment of the mesh allows for large computational savings \cite{rodolfo, liu-dual} since the operations are by far cheaper when integrating the momentum field on coarser meshes, as compared to the single scalar Cahn-Hillard equation without any elliptic equation. 

The multi-resolution method that decouples $\bf u$ and $C$ works as follows. $\bf u$ is projected from a base mesh, which is non-uniform, to a refined uniform mesh on which $C$ resides. The projection employs a tri-cubic Hermite spline interpolation, with a stencil of four points in each direction, for a total of sixty-four points in three dimensions. Here, the Hermitian interpolation is a preferred since the accuracy has been proven to be sufficient for turbulent flows, and is considerably cheaper than other methods such as B-splines \cite{rodolfo}. This stencil is generated only once at the start of the simulation and is reused throughout. To preserve the solenoidal properties of the momentum field, instead of directly projecting $\bf u$, the normal velocity gradients on the base mesh are first computed and then the projection is applied on the normal velocity gradients. Finally, with a refined 2D velocity field interpolated at a reference location (in each direction), the refined velocities are integrated for the entire domain using the interpolated gradients. For the back-coupling of the $C$ field, the refined uniform mesh is directly projected to the stretched mesh since there is no solenoidal requirement for $C$. This down-sampling projection step is used to obtain $\mu$, $\rho$ and ${\bf F}_{st}$. The present method is an improvement over the previous method used in \cite{rodolfo}, since here, the stretched mesh can contain an arbitrary number of nodes employing different stretching parameters. 


\section{Results and discussion}
\label{sec-case}
In Section \ref{sec-shear}, we test the convergence of the results with mesh refinement and the performance of the new discretization scheme for the biharmonic term. Section \ref{sec-buble} shows the ability of the present approach to deal with large density and viscosity contrasts. In Section \ref{sec-rb0}, a possible application of multiphase turbulence is simulated --- Rayleigh-B\'enard convection with drops, where the performance of the multi-resolution meshes is also tested.

\subsection{Drop deformation in shear flow}
\label{sec-shear}

\begin{figure}
\centering
\includegraphics[width=0.6\linewidth]{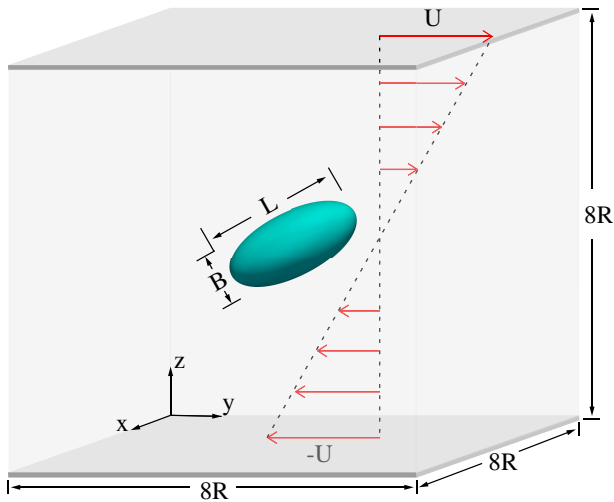}%
\caption{\label{fig-shear1}   Configuration for drop deformation in shear flow.}
\end{figure}

In order to test the mesh refinement convergence of our approach and the performance of the new discretization scheme for the biharmonic term, we consider the deformation of a drop in a shear flow with matched density and viscosity. A drop of radius $R$ is initially placed at the center of a domain of $8R \times 8R \times 8R$, as shown in Fig.~\ref{fig-shear1}. In the domain, there are two no-slip plates moving at a speed of $U$ in opposite direction, and periodic boundary conditions are used in the other directions. Due to the shear stress exerted by the surrounding fluid, the drop elongates until the surface tension counteracts the resulting load. We define the deformation ratio $\Gamma = (L - B)/(L + B)$ as in \cite{dual, shear1, shear2} to quantify the degree of drop deformation, where $B$ and $L$ are the lengths of the minor and major axes of the deformed drop at equilibrium, respectively, see Fig.~\ref{fig-shear1}. The governing dimensionless parameters are the capillary number $\Ca = \mu\dot{\gamma} R/\sigma$, the Reynolds number $\Re= \rho\dot{\gamma} R^2/\mu$, and the Weber number $\We=\rho(\dot{\gamma} R)^2 R/\sigma=\Ca \, \Re$, where $\dot{\gamma}  =2U/H$ is the shear rate and H the thickness of the fluid layer. Gravity is not considered here. With $\Ca \ll 1$ and $\Re \ll 1$, $\Gamma$ is expected to linearly depend on $\Ca$ accounting to  $\Gamma\approx (35/32)\Ca$ \cite{shear0}. 

\begin{figure}
\centering
\includegraphics[width=0.4\linewidth]{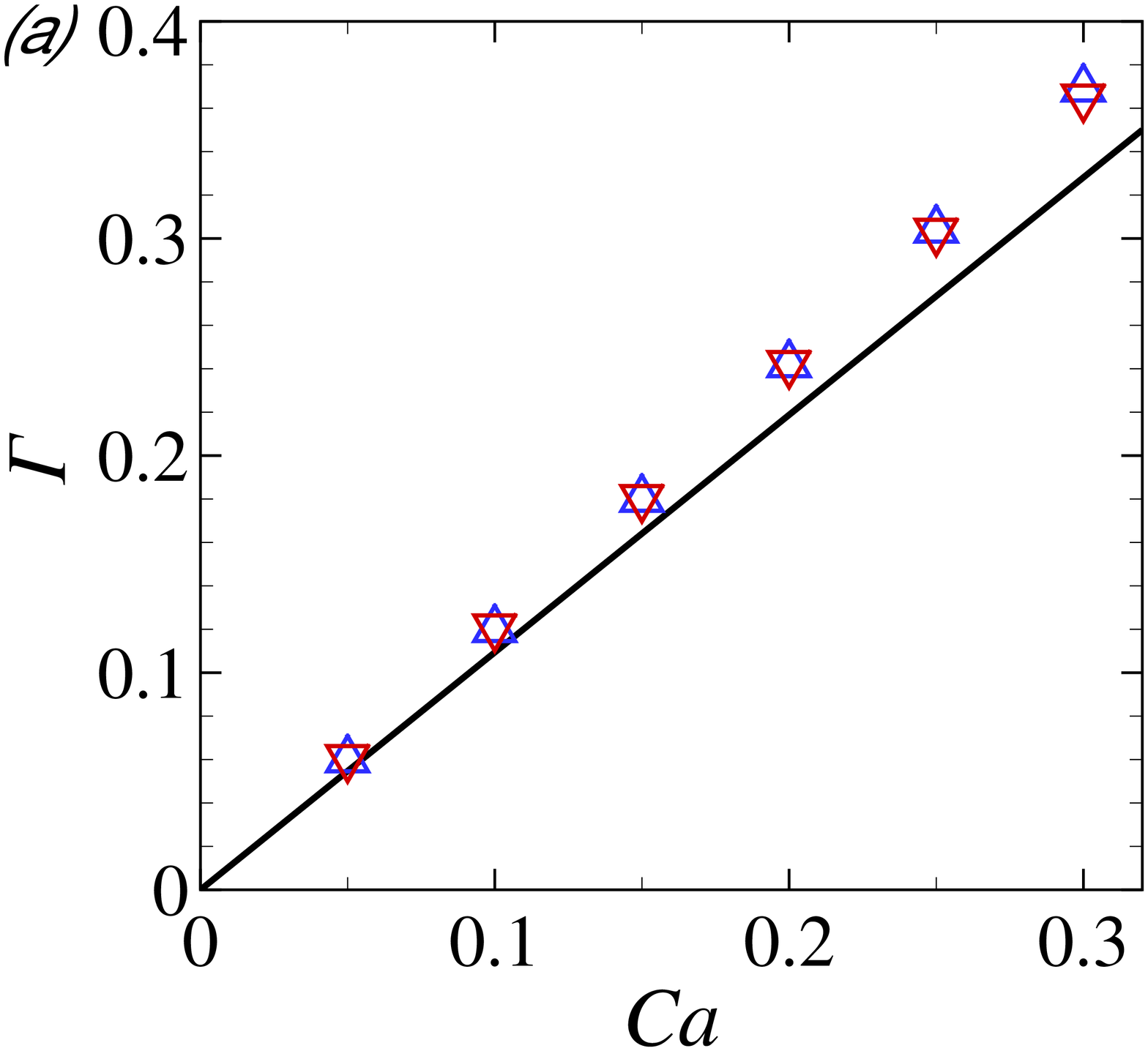}%
\hspace{0.1\linewidth}
\includegraphics[width=0.4\linewidth]{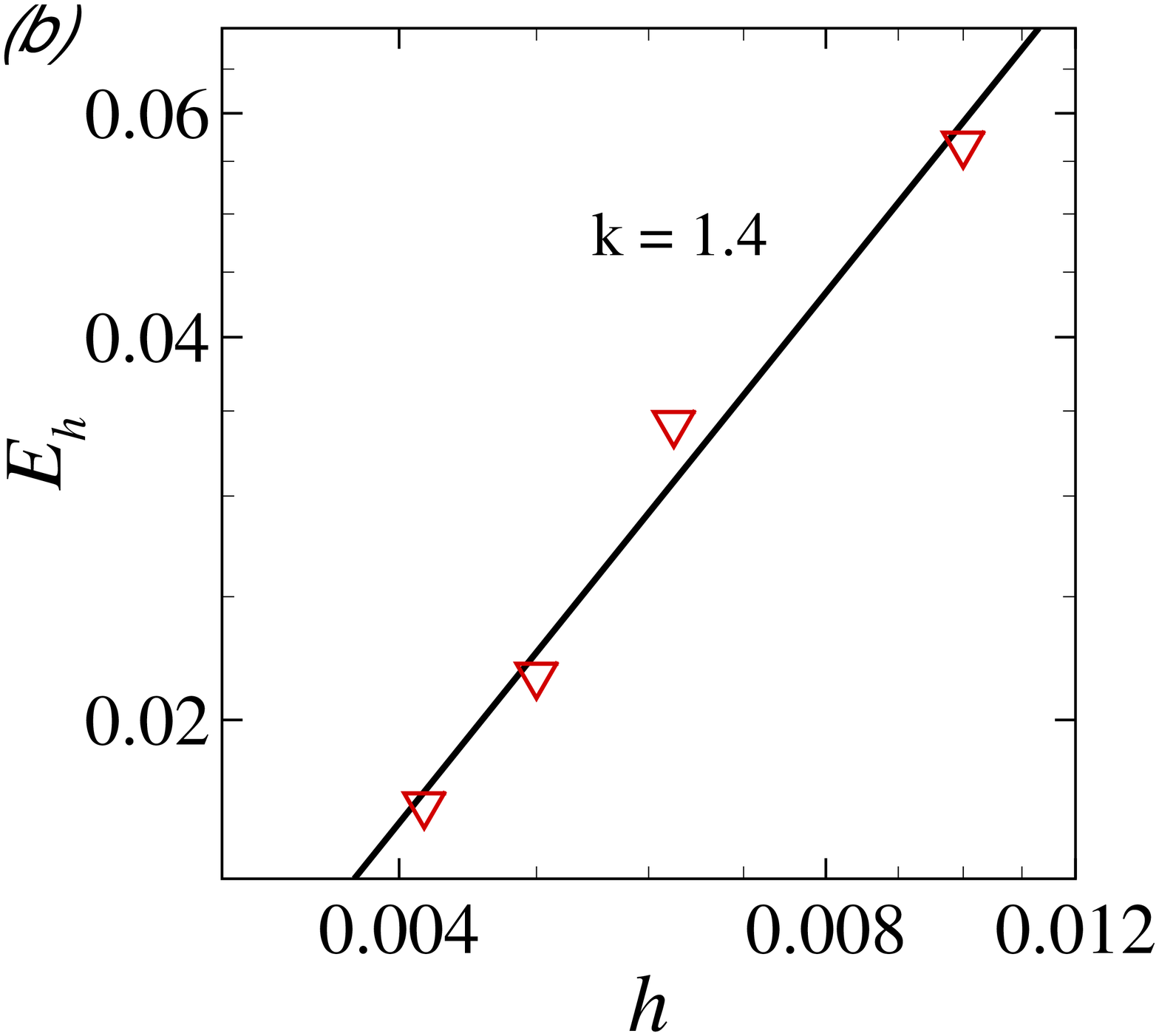}%
\caption{\label{fig-shear2} (a) Comparison of the present results ($\nabla$) with the theoretical approach in \cite{shear0} (black line) and the previous numerical results in \cite{dual} ($\Delta$) in terms of the drop deformation ratio $\Gamma$ at various capillary numbers $\Ca$. (b) Convergence study with mesh refinement at $\Ca=0.1$ in terms of the error $E_h$ of $\Gamma$. The slope of the solid line is $k=1.4$.}

\end{figure}

\begin{figure}
\centering
\includegraphics[width=0.7\linewidth]{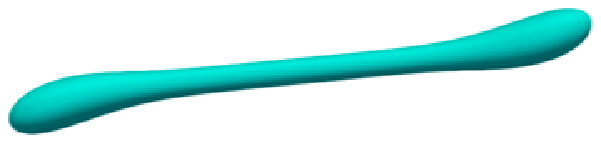}%
\hspace{0pt}
\includegraphics[width=0.7\linewidth]{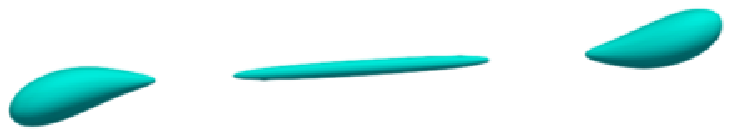}%
\caption{\label{fig-shear3} Breakup of a spherical drop in shear flow at $\Ca=0.39$ and $\Re=1$. The snapshots are at $t=20$ (upper) and $t=29$ (lower).}
\end{figure}

Fig.~\ref{fig-shear2} shows the variation of the deformation ratio $\Gamma$ as function of $\Ca$ at $\Re=0.03$, for simulations performed on a grid with $h=0.005$. The comparison with the theoretical prediction \cite{shear0} and the previous numerical results \cite{dual} gives good agreement. With increasing $\Ca$, the deformation ratio $\Gamma$ becomes larger than the theoretical prediction \cite{shear0} since the assumption of $\Ca \ll 1$ for this prediction is no longer satisfied. As reported in the previous studies \cite{dual, shear1, shear2}, the drop breaks up at $\Ca = 0.39$ and $\Re=1$. We also perform this case in a domain of $12R \times 8R \times 8R$ as shown in Fig.~\ref{fig-shear3}. The drop breaks up into three smaller ones as expected.

\begin{figure}
\centering
\includegraphics[width=0.45\linewidth]{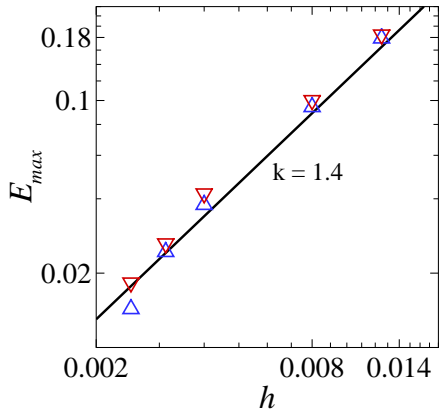}%
\caption{\label{fig-shear4} Convergence study with mesh refinement in terms of $E_{max}$. Red symbols $\nabla$ represents the data with the explicit scheme of Eq.~(\ref{eq-disbi}) and blue $\delta t=5 \times 10^{-5}$, and $\Delta$ with the new implicit scheme of Eq.~(\ref{eq-new}) and $\delta t=2 \times 10^{-3}$. The slope of the solid line is $k=1.4$.}
\end{figure}

Fig.~\ref{fig-shear4} shows the results of the convergence study with different mesh size $h=0.0031$, $0.0042$, $0.0050$, $0.0063$ and $0.0100$ at $\Ca=0.1$. The numerical error $E_h$ is calculated by comparing $\Gamma$ to the value obtained with the finest mesh ($h=0.0031$). The convergence rate is of $1.4$, which is between $1$ and $2$, as expected since the phase-field method \cite{jacqmin00, ding07jcp} for the interface used here is first-order accurate while the NS solver is second order \cite{jcp96}.

We have also tested the performance of the explicit discretization scheme in Eq.~(\ref{eq-disbi}) and the new implicit scheme in Eq.~(\ref{eq-a3d}) for the biharmonic term $\Cn^2\nabla^4 C$ described in Section \ref{sec-bi}. Since the explicit scheme requires a small time step, here we consider the quantity $\Gamma_{max}$, which is reached around $t=0.4$, instead of $\Gamma$ at equilibrium, which is attained only around $t=30$. Here we show a convergence study with mesh refinement at $\delta t = 5\times 10^{-5}$ (the largest value to maintain numerical stability) with the explicit scheme and $\delta t = 2\times 10^{-3}$ with the new scheme in Fig.~\ref{fig-shear4}, where the results agree well. It shows the new implicit scheme in Eq.~(\ref{eq-a3d}) is highly efficient and accurately discretizes the biharmonic term. Thanks to this, we can perform the large-scale simulations of turbulent multiphase flows.

\subsection{Rising bubble with buoyancy}
\label{sec-buble}

\begin{figure}
\centering
\includegraphics[width=0.6\linewidth]{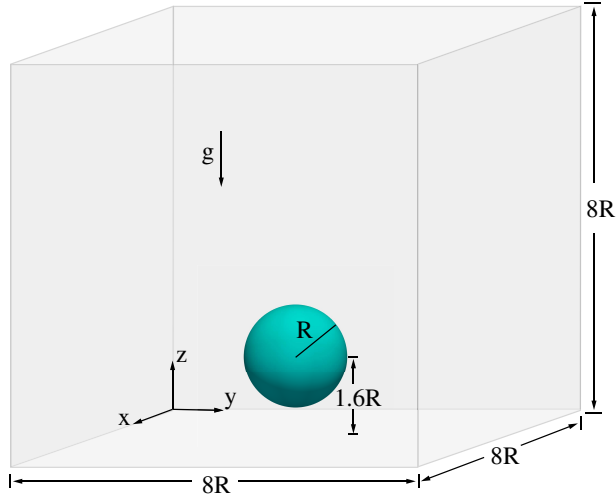}%
\caption{\label{fig-bubble1}   Configuration for rising bubble with buoyancy.}
\end{figure}

In this subsection, we test the performance of the present approach by simulating a three-dimensional bubble rising in liquid water with a large density and viscosity contrast up to $1000$ and $100$ times, respectively, which has the same configuration as previous axisymmetric studies \cite{ding07jcp,rising}. Initially, we place a bubble (fluid 2) of radius $R$ in the domain of $8R \times 8R \times 8R$ with the distance from the bottom plate to bubble center of $1.6R$, as shown in Fig.~\ref{fig-bubble1}. No-slip and non-penetration boundary conditions are enforced at all boundaries. The dimensionless parameters controlling this problem are the Reynolds number $\Re=\rho U R/\mu=100$, the Bond number $\Bo=\rho U R^2/\sigma=200$, and the density and viscosity ratios $\lambda_\rho=0.001$ and $\lambda_\mu=0.01$, respectively. Note that $\Fr=1$ and $\We=\Bo$ due to the characteristic velocity $U=\sqrt{gR}$. The mesh used here is $400 \times 400 \times 400$, where the mesh size is the same as in the axisymmetric simulations \cite{ding07jcp,rising}.

\begin{figure}
\centering
\includegraphics[width=\linewidth]{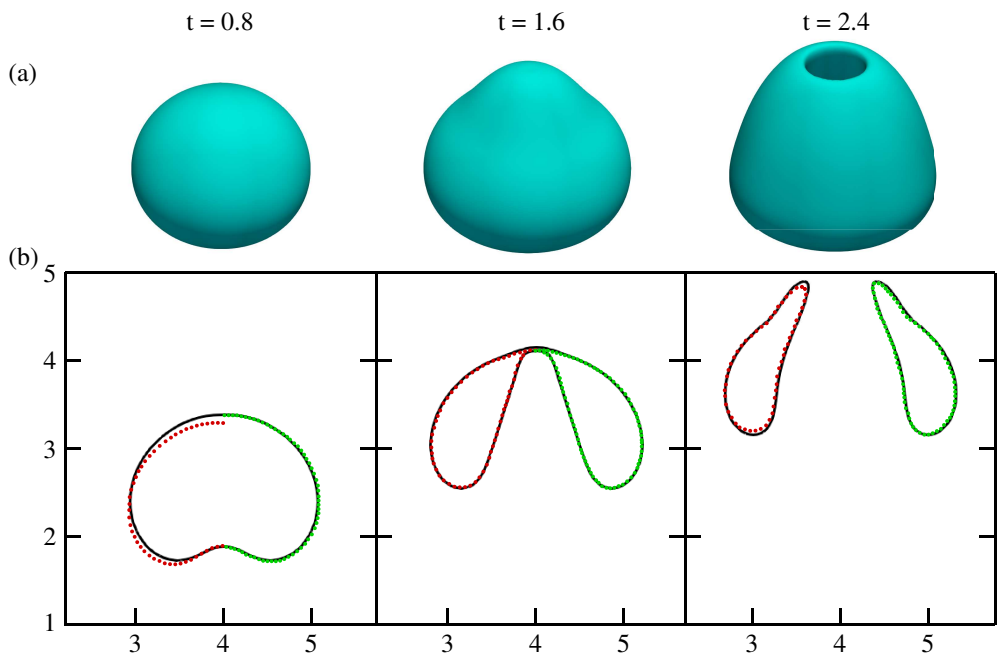}%
\caption{\label{fig-bubble2}  (a) Interface shape of the rising bubble of the present study. (b) Comparison of the results in the $x-z$ plane of the present study (black) with those of the previous studies \cite{ding07jcp} (right half, green dashed line) and \cite{rising} (left half, red dashed line) at $t=0.8$, $1.6$ and $2.4$, respectively.}
\end{figure}

Thanks to buoyancy, the bubble rises. For $\Bo \gg 1$, with the surface tension not large enough to counteract buoyancy, the bottom of the bubble will rise faster than the top, as shown in Fig.~\ref{fig-bubble2}. Therefore, eventually, the bubble breaks up from the tip and evolves into a toroid. Although here a three-dimensional case is performed to test the performance of our code, the flow is indeed axisymmetric, so that we can compare our results with the previous studies of axisymmetric simulations \cite{ding07jcp,rising}. In our numerical simulations, the breakup occurs at t = 1.61 and y = 4.1R, which agrees well with the simulations from previous studies using different numerical approaches, which are t = 1.60 and y = 4.05R with level set method \cite{rising}, and t = 1.61 and y = 4.09R with diffuse-interface method \cite{ding07jcp}. Besides, Fig.~\ref{fig-bubble2} presents the comparisons of the bubble shape at different time instants $t=0.8$, $1.6$ and $2.4$. It shows that also the shape of the bubble's interface in the present study is in good agreement with the previous ones \cite{ding07jcp,rising}.

\subsection{Multiphase turbulent Rayleigh-B\'enard convection}
\label{sec-rb0}

\begin{figure}
\centering
\includegraphics[width=0.6\linewidth]{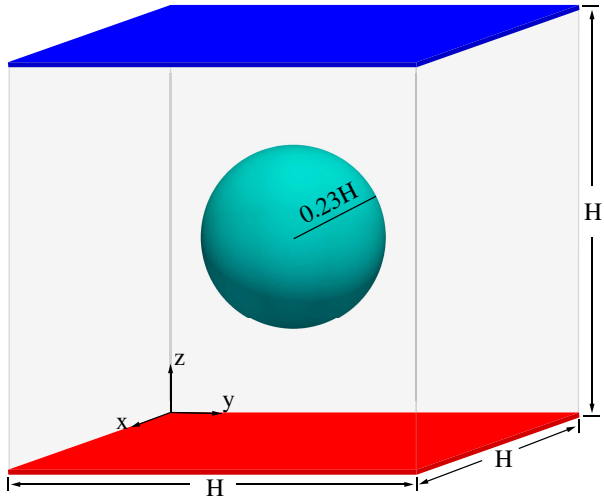}%
\caption{\label{fig-rb1} Configuration for breakup of one big drop in turbulent Rayleigh-B\'enard convection in a domain with a hot bottom plate ($\theta=1$) and a cold top plat ($\theta=0$).}
\end{figure}

Here we consider a possible application of multiphase turbulent flows by using the present approach: Turbulent Rayleigh-B\'enard convection with drops, as shown in Fig.~\ref{fig-rb1}. Rayleigh-B\'enard convection is the motion of a fluid in a cell heated from below and cooled from above \cite{rev1,rev2,chilla}. 

For Rayleigh-B\'enard convection, the temperature advection equation reads
\begin{equation}
\rho c_p \left(\frac{\partial {\theta}}{\partial t} + {\bf u} \cdot \nabla \theta \right)=   \frac{1}{\sqrt {\Pr \Ra }}\nabla \cdot (k_d\nabla \theta),
\label{eq-t}
\end{equation}
where $c_p=k_d/(\kappa \rho)$ is the specific heat capacity. The thermal conductivity $k_d$ is defined as
\begin{equation}
k_d =C + \lambda_{k_d}(1-C),
\label{eq-k}
\end{equation}
where $\lambda_{k_d}=k_{d2}/k_{d1}$ is the ratio of the thermal conductivity. We choose the distance between the hot and cold plates as the characteristic length, and the free fall velocity $U=\sqrt{\alpha_1 g L \Delta}$ as the characteristic velocity. The relevant dimensionless groups of the configuration are the Rayleigh number $\Ra=\alpha_1 \rho_1 g L^3 \Delta/(\mu_1 \kappa_1)$, the Prandtl number $\Pr=\mu_1/(\rho_1 \kappa_1)$,  where $\alpha$ is the thermal expansion coefficient, $\Delta$ the temperature difference and $\kappa$ the thermal diffusivity, in addition to the dimensionless numbers controlling the droplets.

For this case the gravity force ${\bf G}$ in Eq.~\ref{eq-ns} depends on both $C$ and the dimensionless temperature $\theta$, whose effects on density are considered within the Boussinesq approximation,
\begin{equation}
 {\bf G}=\left\{[C+\lambda_\alpha \lambda_\rho (1-C)] \, \theta-\frac{\rho}{\Fr}\right\} {\bf j}, 
\label{eq-g}
\end{equation}
where $\lambda_\alpha=\alpha_2/\alpha_1$ is the ratio of the thermal expansion coefficients $\alpha$. 

\subsubsection{Breakup of one big drop in turbulent Rayleigh-B\'enard convection}
\label{sec-rb}

Initially, a drop of radius $0.23H$ (represented by $C=1$) with matched density and viscosity with the ambient fluid is placed at the center of the domain $H \times H \times H$, with a linear temperature profile and zero velocity. The boundary conditions at the top and bottom plates are set as $C=0$, no-slip condition and fixed temperature $\theta=0$ (top) and $1$ (bottom). Periodic boundary conditions are used in the horizontal directions.  The chosen dimensionless parameters are $\Ra=10^8$, $\Pr=1$ and $\We=8000$. Note that $\We$ here is large because it is defined by the system height instead of the droplet size. For local Weber number which is defined using the droplet size, we find that the value is $O(1)$ after the droplet breakup, which is consistent with the Kolmogorov-Hinze theory \cite{kol,hinze}. The chosen Rayleigh number is large enough for the flow to enter the turbulent regime. The mesh is $500 \times 500 \times 500$, which is consistent with the grid resolution checks in \cite{zhou}. 

\begin{figure}
\centering
\includegraphics[width=\linewidth]{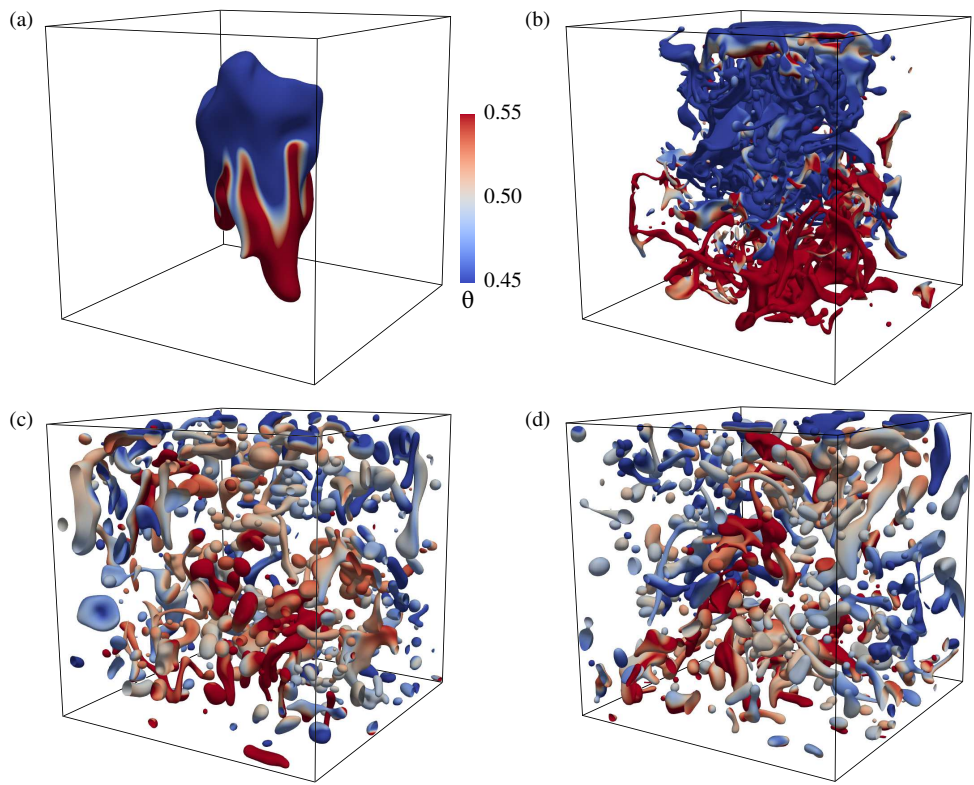}%
\caption{\label{fig-rb2} Snapshots of the interface shape of drops at $\Ra=10^8$, $\Pr=1$ and $\We=8000$. The times are (a) $t=9$, (b) $t=12$, (c) $t=15$ and (d) $t=100$. Temperature on the surface is shown in different colors (hot in red and cold in blue).}
\end{figure}

\begin{figure}
\centering
\includegraphics[width=0.45\linewidth]{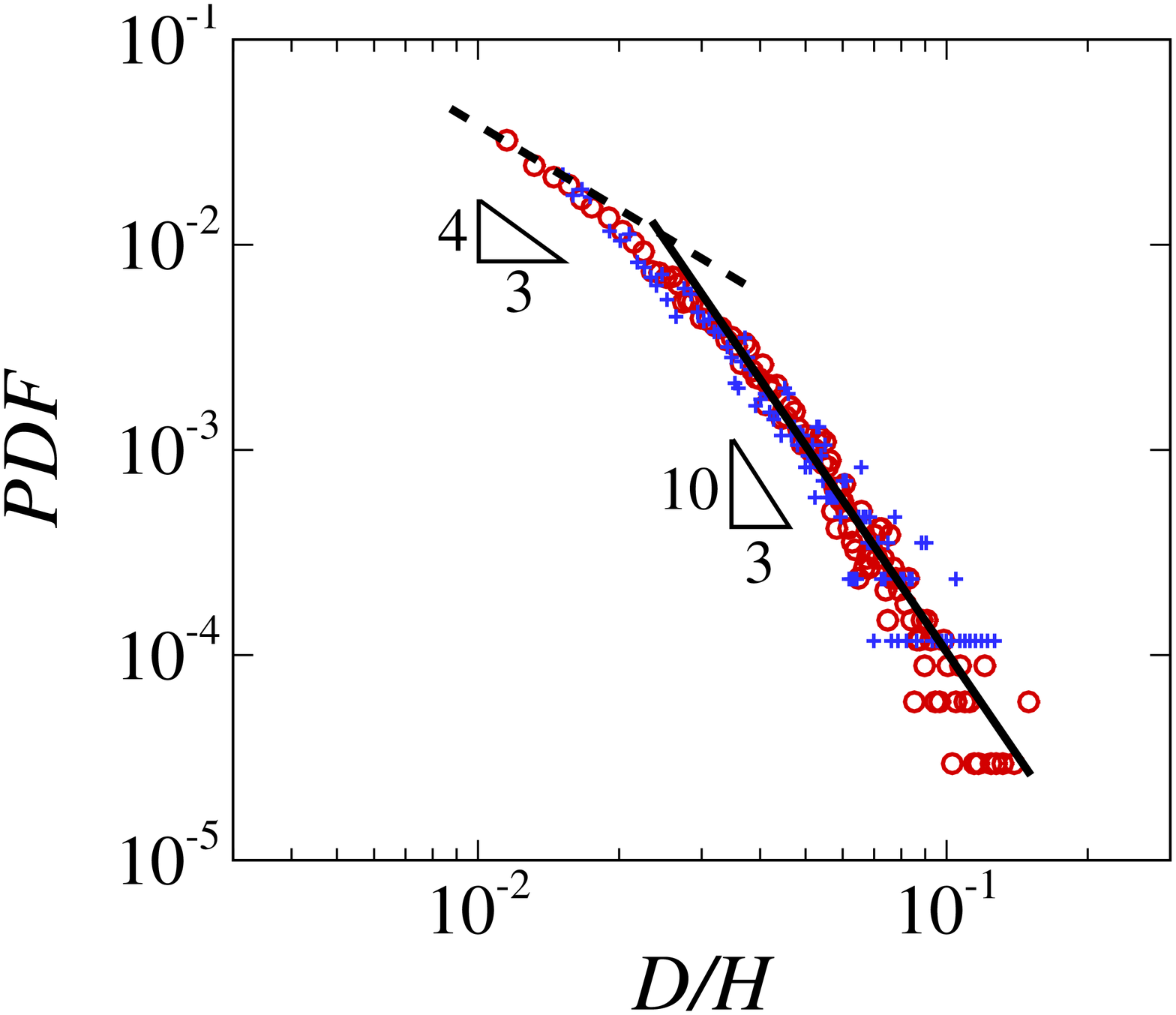}%
\caption{\label{fig-rb3} Probability distribution function (PDF) of the drop size $D/H$ calculated from the drop volume. The solid and dashed lines indicate the scaling laws $-10/3$ \cite{pdf} and $-4/3$ \cite{pdfjfm}, respectively. The red circles denote the results on the single resolution meshes, and the blue cross symbols on the multi-resolution meshes.}
\end{figure}

Fig.~\ref{fig-rb2} shows snapshots of the drops in Rayleigh-B\'enard convection. The drops first deform due to buoyancy (see Fig.~\ref{fig-rb2}a), and then breaks up because of the small surface tension (see Fig.~\ref{fig-rb2}b). As time evolves, hundreds of drops of various sizes are advected in the turbulent field (Fig.~\ref{fig-rb2}c and \ref{fig-rb2}d). The drop size is characterized by an effective diameter $D$, which is defined as $\frac{4\pi}{3} (D/2)^3=V$, with $V$ being the drop volume. The resulting distribution of the drop sizes is shown in Fig.~\ref{fig-rb3}. We observe that the probability distribution function (PDF) of the large drops follows the scaling $(D/H)^{-10/3}$ while that of the small drops obeys the scaling $(D/H)^{-4/3}$, which both originate from the previous theory studies for the respective regimes \cite{pdf,pdfjfm}: First, in turbulent flows, the distribution of the drop size has been studied extensively. The well-known $-10/3$ scaling law for the large drops in turbulence was proposed in ref. \cite{pdf} and validated by many experimental and numerical studies \cite{nature02,pipe,pop16,LB19JFM}. Second, the derivation of the $-4/3$ scaling law for the relatively small drops originates from a recent study \cite{pdfjfm}. It is based on the energy balance in a regime dominated by surface tension. Fig.~\ref{fig-rb3} shows that the present numerical simulations and the theoretical analyses \cite{pdf,pdfjfm} give consistent results.

\begin{figure}
\centering
\includegraphics[width=0.45\linewidth]{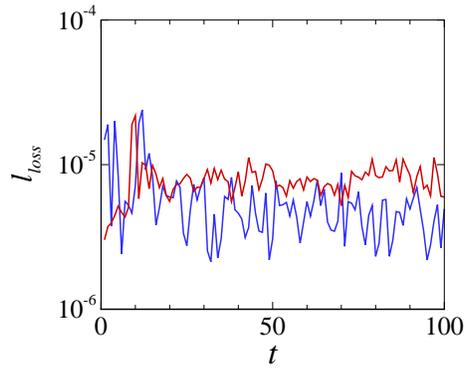}%
\caption{\label{fig-rb4} Time evolution of the mass loss $l_{loss}$ of drops in turbulent Rayleigh-B\'enard convection. The blue and red curves represent the results on the single and multi-resolution meshes, respectively.}
\end{figure}

The mass conservation is also tested in this section. Fig.~\ref{fig-rb4} shows the normalized mass loss $l_{loss}= (m_t-m_0)/m_0$, where $m_t$ is the mass of fluid $1$ (drops) at time $t$ and $m_0$ is the initial mass of the drop. We see that the maximal mass loss is of the order of $10^{-5}$ and the value of $l_{loss}$ is not increasing in time. This demonstrates the good mass conservation in the present approach, which is consistent with the other studies with phase-field methods \cite{ding07jcp, shu03, soldati3}.

\begin{figure}
\centering
\includegraphics[width=0.4\linewidth]{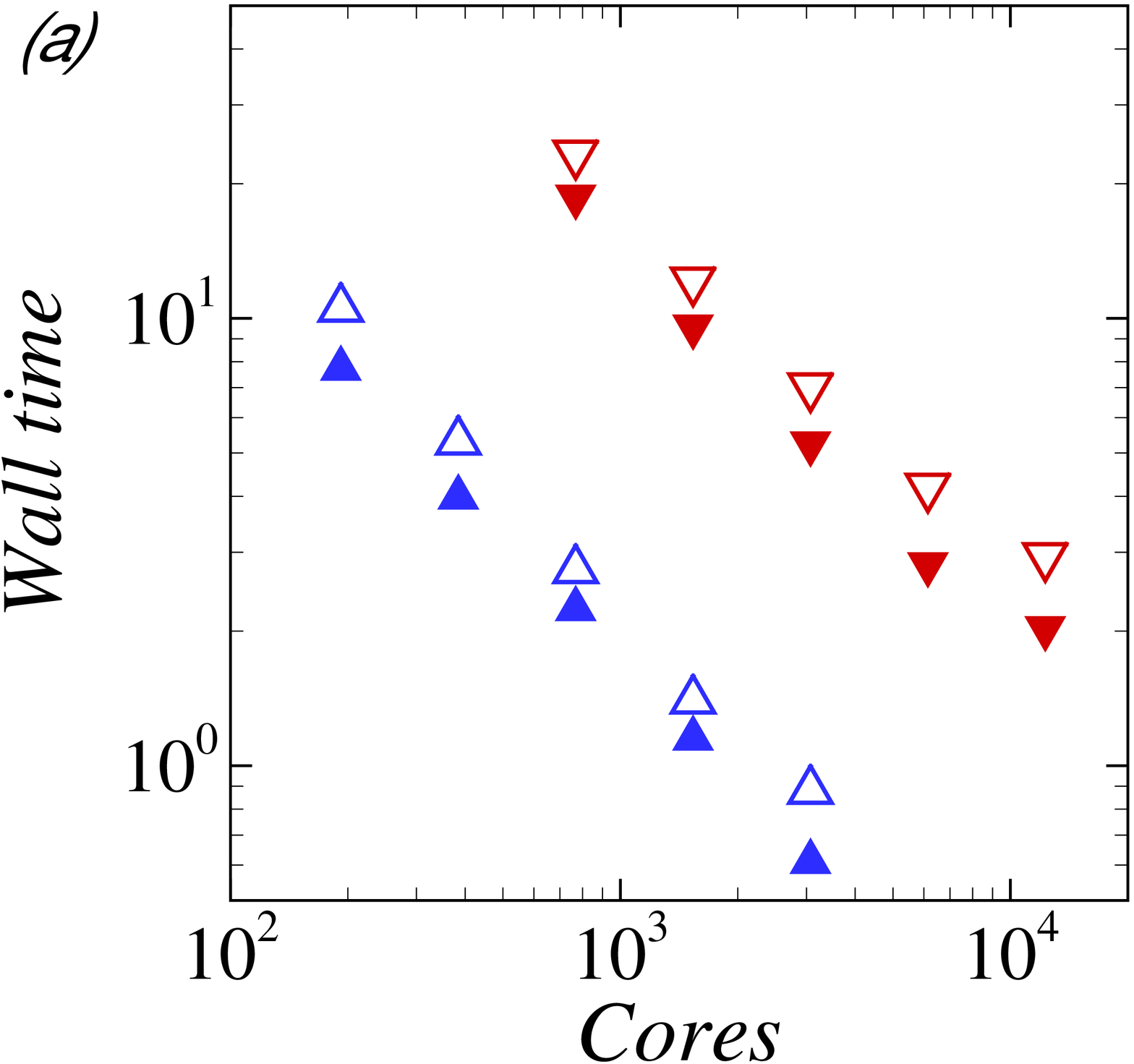}%
\hspace{0.1\linewidth}
\includegraphics[width=0.4\linewidth]{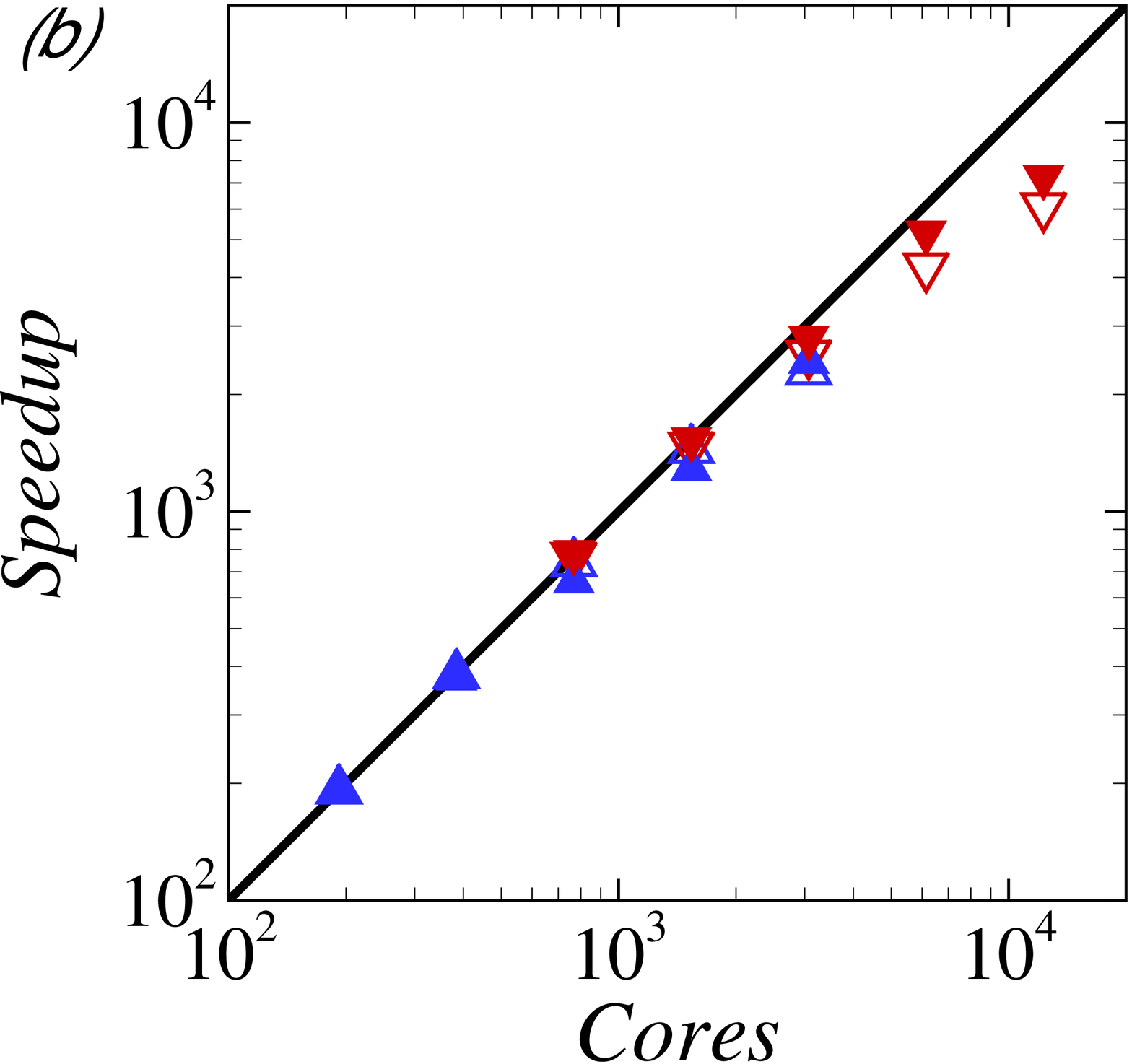}%
\caption{\label{fig-rb5} (a) Wall time and (b) speedup of the computation time compared to that with single core as functions of CPU cores with gridpoints of $1000^3$ ($\Delta$) and $2000^3$ ($\nabla$). The empty symbols are the present data, and the filled symbols the data of turbulent single-phase flows \cite{gpu}.}
\end{figure}

We also simulated the case on multi-resolution meshes with otherwise unchanged parameters, uniform mesh of $500^3$ for the CH equation and stretched mesh of $250^3$ for the NS equation, i.e. the same resolution for volume fraction $C$ and a coarser one for velocity $\bf u$ and temperature $\theta$ compared to the single-resolution gird. The consistent results obtained on the multi- and single-resolution meshes are shown in Fig.~\ref{fig-rb3} and \ref{fig-rb4} in terms of PDF of $D/H$ and the time evolution of $l_{loss}$.

We also test the computational efficiency of the method on the supercomputer MareNostrum at the Barcelona Computing Center
(2 sockets Intel Xeon Platinum 8160 CPU with 24 cores each @ 2.10GHz, for a total of 48 cores per node). Two sets of gridpoints are used, i.e. $1000^3$ and $2000^3$, and the option of multi-resolution is not used here to fit the setting of the previous study. The wall clock time per step and the speedup comparing with a single core as functions of CPU cores are presented in Fig.~\ref{fig-rb5}. Compared to the AFiD code for single phase flows \cite{gpu}, the computational cost of the present approach for the multiphase flows is only less than $1.5$ times more. Moreover, the parallel efficiency is quite good until the CPU cores used are more than $3072$. These data show that the computational performance of the present approach for turbulent multiphase flows is nearly as good as the solver for turbulent single-phase flows.

\subsubsection{Coalescence of $O(10^3)$ drops in Rayleigh-B\'enard convection}
\label{sec-1000}

\begin{figure}
\centering
\includegraphics[width=0.7\linewidth]{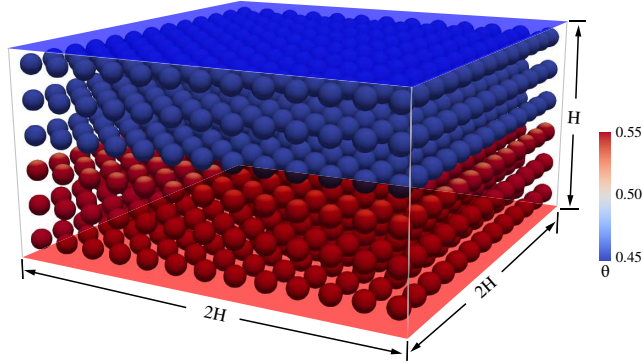}%
\caption{\label{fig-1000i} Initial configuration for the study of coalescence of $O(10^3)$ drops in turbulent Rayleigh-B\'enard convection. The color code represents the temperature.}
\end{figure}

\begin{figure}
\centering
\includegraphics[width=\linewidth]{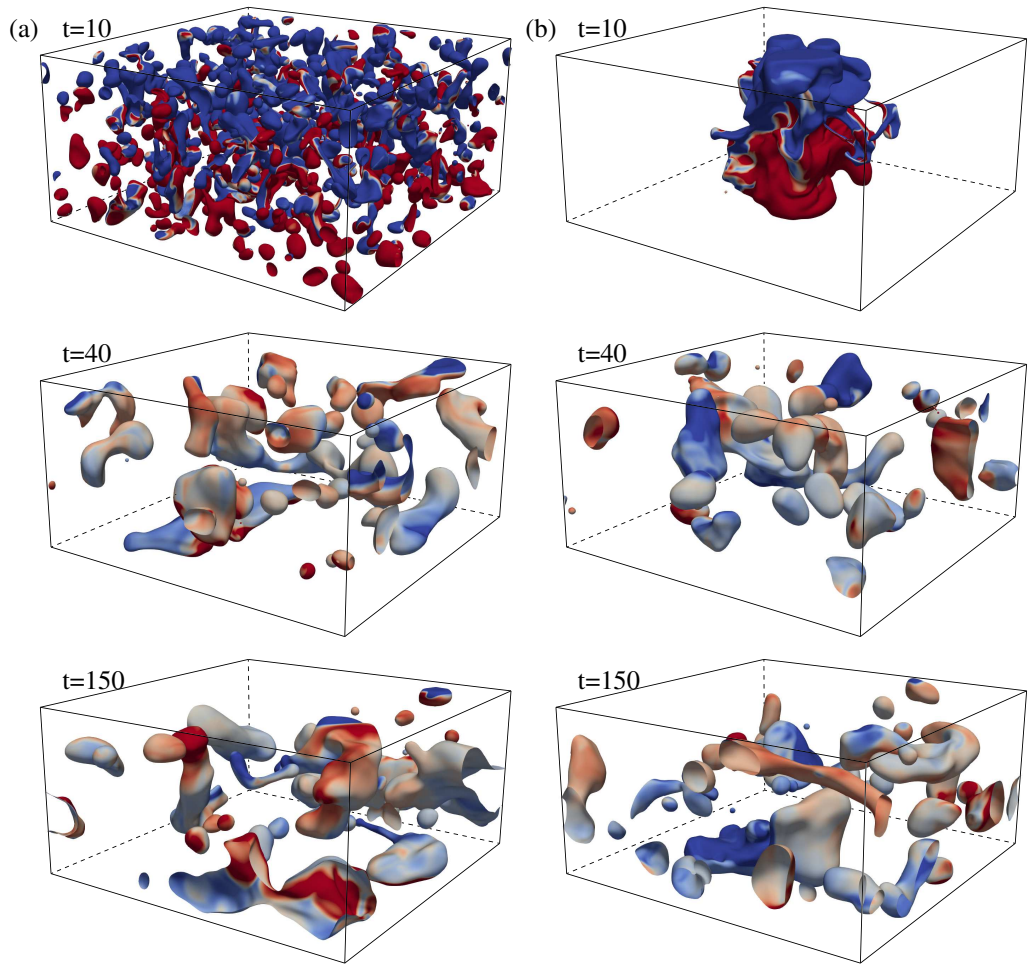}%
\caption{\label{fig-1000t} Snapshots of the interface shape of drops at $\Ra=10^8$, $\Pr=1$ and $\We=1000$ with initially (a) $1000$ drops and (b) only one drop. Temperature on the surface is shown in the same color bar of Fig.~\ref{fig-rb2}. }
\end{figure}

The topological change of the interface includes the breakup and coalescence of drops. In Section \ref{sec-rb}, we clearly observed the breakup of drops. In this section, we will show the coalescence of $O(10^3)$ drops in turbulent Rayleigh-B\'enard convection. The initial setup is presented in Fig.~\ref{fig-1000i}, where we placed $1014$ drops with a uniform diameter of $0.08H$ in a domain of $2H\times 2H \times H$. The simulation was performed on the mesh of $1000\times 1000\times500$ and $2048$ CPU cores. The Weber number was set to $\We=1000$, which is smaller than that in Section \ref{sec-rb}. The other dimensionless parameters and boundary conditions are the same as in Section \ref{sec-rb}. 

As seen from the snapshots at $t=10$, $40$ and $150$ in Fig.~\ref{fig-1000t} (a), most of drops coalesce into larger ones. Since the Weber number here is smaller than that in Section \ref{sec-rb}, surface tension here is stronger and can resist inertia, leading to larger drop sizes. 

We also simulated a case with a different initialization, where only one big drop with a diameter of $0.8H$ is placed at the center of the domain. Although different initial conditions are used, similar statistic equilibrium states were obtained after sufficiently long times (see Fig.~\ref{fig-1000t}).

\section{Conclusion}
\label{sec-con}
In this study we have shown how to efficiently implement the phase-field method into the single-phase DNS solver AFiD. A new discretization scheme for the biharmonic term $\Cn^2\nabla^4 C$ of the Cahn-Hilliard equation has been proposed. Together with the approximate-factorization method, the FFT-based Poisson solver, and a pencil distributed parallel strategy, massive DNSs (up to 8 billion gridpoints and 3072 CPU cores are used) for turbulent multiphase flows can be performed. 

The suggested new approach has then been validated by comparisons with several numerical experiments. In the case of drop deformation in shear flow, the results agree well with theoretical and previous numerical results, and the convergence study with mesh refinement shows an accuracy between first and second order, as expected. Then, also for the case of a rising bubble with buoyancy, good agreement is achieved when comparing our results with previous simulations, even with large density or viscosity contrast of up to $1000$ or $100$ times, respectively. Furthermore, in the case of breakup and coalescence of drops in turbulent Rayleigh-B\'enard convection, we observe good performance of our approach to deal with turbulent multiphase flows, including good mass conservation and high efficiency of computation, thus establishing our scheme to perform reliable simulations for turbulent multiphase flows in large-scale computations.

The new scheme and code therefore offer great opportunities to better understand the physics of turbulent two-phase flow with coalescence and breakup of droplets and bubbles.

\section*{Acknowledgments}
This work was financially supported by ERC-Advanced Grant under the project no. 740479. We acknowledge PRACE for awarding us access to MareNostrum in Spain at the Barcelona Computing Center (BSC) under the project 2020225335, and Irene at Tr\'es Grand Centre de calcul du CEA (TGCC) under the project 2019215098. This work was also carried out on the national e-infrastructure of SURFsara, a subsidiary of SURF cooperation, the collaborative ICT organization for Dutch education and research.



 \bibliographystyle{model1-num-names}


\end{document}